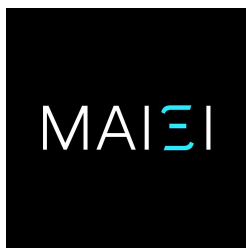

# Montreal AI Ethics Institute

*An international, non-profit research institute helping humanity define its place in a world increasingly driven and characterized by algorithms*

**Website**: https://montrealethics.ai
**Newsletter**: https://aiethics.substack.com

# Report prepared by the Montreal AI Ethics Institute (MAIEI) on Publication Norms for Responsible AI

*Based on insights and analysis by the Montreal AI Ethics Institute (MAIEI) staff and supplemented by workshop contributions from the AI Ethics community co-hosted by MAIEI & Partnership on AI on May 13th 2020 and May 20th 2020*

Primary contacts for the report:

Abhishek Gupta, Founder and Principal Researcher, MAIEI
Machine Learning Engineer and CSE Responsible AI Board Member, Microsoft
abhishek@montrealethics.ai

Camylle Lanteigne, AI Ethics Researcher and Research Manager, MAIEI
Ethics Analyst, Algora Lab
camylle@montrealethics.ai

Victoria Heath, AI Ethics Researcher, MAIEI
victoria@montrealethics.ai

## Overview of Our Recommendations

- **Create tools to navigate publication decisions:** The use of extrinsic measurements like benchmarks, or third-party expert panels could be a crucial step to navigating publication decisions in a fair way. Such methods could set a certain standard in terms of the acceptable level of risk associated with a publication. In line with this suggestion, it would also be pertinent to keep a record of the papers that were rejected due to their inherent risk, as well as some metrics on these.
- **Offer a page number extension:** It may be beneficial to extend the page limit for published papers to allow researchers to include negative results (results that are insignificant or disprove researchers' hypotheses), which aren't traditionally printed. In addition to expanding the number of pages, there should also be a significant change in the culture surrounding the publication of negative results.
- **Develop a network of peers:** Developing a network of peers to evaluate researchers' AI models in terms of potential risks and benefits may be an important tool towards better and safer publication norms. If such a mechanism were put in place, the evaluation could be fully or partly based on philosopher John Rawls's "veil of ignorance." If this idea was applied to reviewing AI research, peer reviewers could be asked to consider the potential advantages and risks of a new AI research from different social perspectives.
- **Require broad impact statements:** The NeurIPS conference requires that submitted papers include a statement of broader impact with respect to the research presented. This incentivizes researchers to think about potential risks and benefits by making reflection a requirement for one's work to be considered at NeurIPS. Similar measures at all conferences and publications would encourage researchers to critically assess their research in terms of its effects, positive and negative, on the world
- **Require the publication of expected results:** Requiring that researchers write and publish the expected results of their research project (including but not limited to its broader social and ethical impacts) could help foster reflection around potential benefits and harms even before researchers undertake their project.
- **Revamp the peer-review process:** The well-established practice of peer review is a great opportunity for exchanges on the risks and benefits each reviewer sees in the paper they are revising. If a question or requirement were added to this effect when papers were reviewed, it may have a rapid and widespread impact in inciting researchers to consider what may follow from their research. An effective review process should promote limiting risks while also being clear, fair, and efficient. One way of doing this is to intensify the requirements for publication proportionally to how risky the research is deemed by peers reviewing the paper.

## Introduction: What History Can Teach Us

The history of science and technology shows that seemingly innocuous developments in scientific theories and research have enabled real-world applications with significant negative consequences for humanity; from the eugenics movement in the late 19th and early 20th centuries[1] to the cataclysmic nuclear destruction in Japan in 1945[2]. This history reveals that the real-world impacts of scientific research cannot be separated from the research itself, and it's important to look at the social, cultural, political, and economic realities that shape the way science is used and the norms that regulate it. It's also important to consider the impacts these developments may have on people across the world. Often, researchers do not consult with or consider individuals that may be negatively affected by scientific developments, reflecting existing power imbalances in which a small group of privileged individuals make decisions or take risks that impact millions, if not billions, of lives. We need to be aware of these power imbalances and how they're rooted in existing inequalities—this is especially true in the field of artificial intelligence (AI). The pace and scale of impact from AI far exceed other technologies. Thus, critical examination is necessary, especially as this technology becomes increasingly deployed throughout society.

The aforementioned power imbalances and inequalities in scientific research are apparent in the general disconnect between the research priorities of funders (e.g. grant-making bodies, companies, governments, etc.) and the broader societal interest. The academic paradigm of "*publish or perish*"[3] and the undue pressure it creates, overshadows more fundamental questions that need to be asked. The most important of which is, "Why is this research project being pursued in the first place?" This lack of critical reflection and external pressure has given rise to predatory journals with lax quality standards regarding what gets accepted for publication—this is especially an issue for researchers who are new to an academic field and are uncertain about publishing norms. For example, research in phrenology continues to be accepted in highly revered journals[4] despite decades of precedent demonstrating that phrenology is pseudoscience. The egregious inclusion of this type of research resulted in severe backlash from scholars that led to a retraction and apology from the journal. This example highlights how fallacious research can slip through the cracks, especially without critical reflection.

---

[1] MacKenzie, D. (1978). Statistical theory and social interests: a case-study. *Social studies of science*, 8(1), 35-83. Retrieved from: https://journals.sagepub.com/doi/10.1177/030631277800800102
[2] Bridger, S. (2011). Scientists and the Ethics of Cold War Weapons Research (Doctoral dissertation, Columbia University).
[3] Rawat, S., & Meena, S. (2014). Publish or perish: Where are we heading?. *Journal of research in medical sciences: the official journal of Isfahan University of Medical Sciences*, 19(2), 87.
[4] Venkataramakrishnan, S. (2020, June 24). *Top researchers condemn 'racially biased' face-based crime prediction*. Financial Times. https://www.ft.com/content/aaa9e654-c962-46c7-8dd0-c2b4af932220

In order to ensure that the science and technology of AI are developed in a humane manner, we must develop research publication norms that are informed by our growing understanding of AI's potential threats and use cases. Unfortunately, there haven't been many efforts to create a set of publication norms for responsible AI because the field of AI is currently fragmented in terms of how this technology is researched, developed, funded, etc. Thus, the norms around AI publications and ethical standards are not only fragmented, but also contradictory in many cases. A standardized approach to publication norms in AI across a large number of jurisdictions is essential. Many subfields in AI are experiencing a boom in interest, with growing demands to produce novel research. Thus, there is an elevated risk of a lack of awareness on what adequate and rigorous publishing norms are. Additionally, there is a high degree of susceptibility to predatory journals[5] that lure budding researchers through a "pay-to-play" model. This amplifies the potential impacts of harmful research that must be critically reviewed before published for public consumption.

Most scientific and academic journals have particular guidelines for submissions, which are a form of publication norms. However, problematic research can still be published. For example, researchers exerted external pressure on Springer[6] to withdraw the publication of a paper in which the researchers claimed to "predict criminality" using neural networks for facial recognition. This demonstrates that the publication ecosystem requires norms with multiple filters. As well as editorial boards that have sufficient demographic diversity and a range of expertise to be able to flag problematic research and prevent its publication, especially in cases where there is potential to render harm on marginalized peoples. However, the prevalent practice in peer-reviewed journals is to evaluate a work purely on its scientific merit, which overlooks the inherent interaction between fundamental research and the social context in which it's conducted. Publication norms have a strong role to play in ensuring interdisciplinarity in review processes so that too narrow a focus does not allow potentially harmful work to pass as seemingly innocuous.

Beyond the social implications of research, it's also important to consider the technical implications. For example, in cryptography, it's important to find vulnerabilities in a system to improve robustness. Cryptographers do this by looking at a system from an adversarial perspective, as well as by sharing their systems openly so they can be evaluated for undue risks by as many people as possible. For example, through this process decades ago, unnecessary risks arose from the use of substitution boxes (S-box)[7] in data encryption standard (DES), which led to the subsequent discovery of differential cryptanalysis. This is also important in AI. Disconcertingly, a lot of the models today aren't protected from machine learning security (MLSec) vulnerabilities, such as data poisoning[8], model inversion[9], and model dumping.[10] In addition to adversarial robustness[11] and common MLSec vulnerabilities, machine learning models may suffer from bugs like deserialization vulnerabilities, or may accept untrusted input.[12] Without transparency about algorithms, implementation, and application, many of these vulnerabilities may go undetected[13], or bundled software versions may not be upgraded.[14] This type of vulnerability was seen in the JBoss middleware, which shipped quietly with dozens of consumer grade softwares but often with an unpatched, or even unpatchable, version.[15] Similarly, software which incorporates AI that relies on an unpatchable library will be vulnerable.

---

[5] Shamseer, L., Moher, D., Maduekwe, O., Turner, L., Barbour, V., Burch, R., and Shea, B. J. (2017). Potential predatory and legitimate biomedical journals: can you tell the difference? A cross-sectional comparison. *BMC medicine*, 15(1), 1-14.

[6] Hatmaker, T. (June 2020). *AI researchers condemn predictive crime software, citing racial bias and flawed methods*. TechCrunch. Retrieved from:
https://techcrunch.com/2020/06/23/ai-crime-prediction-open-letter-springer/

[7] Gargiulo, J. (2002). S-Box Modifications and Their Effect in DES-like Encryption Systems. SANS Institute InfoSec Reading Room. Retrieved from:
https://www.sans.org/reading-room/whitepapers/vpns/paper/768

To expose such vulnerabilities, as well as better understand the social implications of AI research, journal editors must be better equipped to identify all actors who may engage with AI research and make well-informed decisions around whether research should be published, and how. We believe developing a standardized approach to publication norms in AI across a large number of jurisdictions is the first step towards ensuring that the science and technology of AI are developed in a humane manner.

**Initial Recommendations**

---

1. **Create tools to navigate publication decisions**

The use of extrinsic measurements like benchmarks, or third-party expert panels could be a crucial step to navigating publication decisions in a fair way. Such methods could set a certain standard in terms of the acceptable level of risk associated with a publication. In line with this suggestion, it would also be pertinent to keep a record of the papers that were rejected due to their inherent risk, as well as relevant metrics. For example, what subfield of AI was the paper part of? What institution or corporation was the authors affiliated with (if any)? Demographic and identity metrics may also be pertinent to help ensure the review process is not discriminatory. To this end, mechanisms like Retraction Watch that are invested in by the scientific community and held to the highest academic standards[16] will be a cornerstone to strategies combating spurious research. Awareness of such mechanisms and their integration into tools like Google Scholar, akin to COVID-19 research warnings for non-peer-reviewed articles, is essential.

---

[8] Alfeld, S., Zhu, X., and Barford, P. (2016, February). Data Poisoning Attacks against Autoregressive Models. In AAAI (pp. 1452-1458). Retrieved from:
https://www.aaai.org/ocs/index.php/AAAI/AAAI16/paper/download/12049/11758
[9] Fredrikson, M., Jha, S., and Ristenpart, T. (2015, October). Model inversion attacks that exploit confidence information and basic countermeasures. In Proceedings of the 22nd ACM SIGSAC Conference on Computer and Communications Security (pp. 1322-1333).
[10] Chakraborty, A., Alam, M., Dey, V., Chattopadhyay, A., and Mukhopadhyay, D. (2018). Adversarial attacks and defences: A survey. arXiv preprint arXiv:1810.00069.
[11] Carlini, N., Athalye, A., Papernot, N., Brendel, W., Rauber, J., Tsipras, D., ... & Kurakin, A. (2019). On evaluating adversarial robustness. arXiv preprint arXiv:1902.06705.
[12] Kegelmeyer, W. P., Wendt, J. D., and Safta, C. (2019). An Overview of Training Data Security Vulnerabilities: Machine Learning is a Leaky Black Box (No. SAND2019-6536PE). Sandia National Lab.(SNL-CA), Livermore, CA (United States); Sandia National Lab.(SNL-NM), Albuquerque, NM (United States).
[13] Gu, T., Dolan-Gavitt, B., and Garg, S. (2017). Badnets: Identifying vulnerabilities in the machine learning model supply chain. arXiv preprint arXiv:1708.06733.
[14] Security Bulletin: A security vulnerability has been identified in TensorFlow shipped with PowerAI. - https://www.ibm.com/blogs/psirt/security-bulletin-a-security-vulnerability-has-been-identified-in-tensorflow-shipped-with-powerai/
[15] Biasini, N. (2016). SamSam: The Doctor Will See You, After He Pays The Ransom. Talos Blog. Retrieved from: https://blog.talosintelligence.com/2016/03/samsam-ransomware.html

### 2. Offer a page number extension

It may be beneficial to extend the page limit for published papers to allow researchers to include negative results (results that are insignificant or disprove researchers' hypotheses), which aren't traditionally printed. There is a bias towards publishing papers with positive results. That is, results that confirm or partially confirm one's hypothesis.[17] In addition to expanding the number of pages, there should also be a significant change in the culture surrounding the publication of negative results. While a higher maximum of pages may help, it should be accompanied by other measures to encourage researchers to share their negative results and push the broader scientific community to consider negative results in their analysis. Publishers could require the inclusion of all negative results that led the researchers to the positive results published and/or submitted. These could also be indexed through a standardized mechanism for retrieval by engines like Google Scholar so that downstream researchers are aware of what areas have already been explored. Improving citations and value of such negative results work will also incentivize the ecosystem to invest in making negative results more public and elevate the quality of scientific publishing in the space.

### 3. Develop a network of peers

Developing a network of peers to evaluate researchers' AI models in terms of potential risks and benefits may be an important tool towards better and safer publication norms. If such a mechanism were put in place, the evaluation could be fully or partly based on philosopher John Rawls's "veil of ignorance," which asks individuals to create a new society and choose its governing principles without knowing their individual characteristics (e.g. gender, race, social class, etc.).[18] The idea is that they'll choose principles that will benefit everyone in society since they're "ignorant" of their personal circumstances and social standing. If this idea was applied to reviewing AI research, peer reviewers could be asked to consider the potential advantages and risks of a new AI research from different social perspectives. Specific details about fictional but realistic situations and personas may be used to help reviewers think more accurately and relevantly about privileges and disadvantages they might not have themselves. Of course, this shouldn't replace important efforts to make the field of AI more diverse and inclusive—but the exercise of imagining how another person may be affected, positively or negatively, by an AI model is relevant for everyone. This practice has precedent in design thinking approaches[19] and has shown to create products and services that are ultimately more empathetic to the needs of the communities that they are meant to serve and encourage the construction of more inclusive work.

---

[16] Teixeira da Silva, J. A. (2016). Letter to the Editor: Evidence of Bias, Opacity and Lack of Reciprocity by Retraction Watch. KOME: AN INTERNATIONAL JOURNAL OF PURE COMMUNICATION INQUIRY, 4(2), 82-85.

[17] Mlinarić, Anna; Horvat, Martina; Šupak Smolčić, Vesna. (2017). Dealing with the positive publication bias: Why you should really publish your negative results. *Biochemia Medica*, *27(3)*, 447-452. Retrieved from: https://www.ncbi.nlm.nih.gov/pmc/articles/PMC5696751/

[18] Huang, Karen; Greene, Joshua D.; Bazerman, Max. (2019). Veil-of-ignorance reasoning favors the greater good. *PNAS, 116 (48),* 23989–23995. Retrieved from: https://www.pnas.org/content/116/48/23989

### 4. Require broad impact statements

An important method to incite researchers to reflect on the impacts of their work is to give them incentives to do so. For example, the NeurIPS conference recently added a requirement for papers that are submitted: they must include a statement of broader impact with respect to the research presented.[20] This incentivizes researchers to think about potential risks and benefits by making reflection a requirement for one's work to be considered at NeurIPS, one of the most prestigious conferences in the field of AI. Therefore, similar measures at all conferences and publications may be a good way of encouraging researchers to critically assess their research in terms of its effects, positive and negative, on the world. ICLR has followed suit and has included a Code of Ethics[21] for their 2021 edition which is an important step in the proliferation of these standards becoming mainstream, not just something that one gives a brief nod to.

It's important to note that *how* the research community adopts the formulation of broader impact statements is of fundamental importance. If broader impact statements become yet another "ethics stamp" on poorly conceived research, then the entire endeavour becomes counterproductive. Researchers need to critically evaluate the broader impacts of their research *before* starting a project and use these insights to guide a conscientious research methodology and not construct these perspectives after the fact. The point of writing a broader impact statement must be to introspect the ethics of one's work and *not as a mechanism to* avoid desk rejection. One way of doing so is to popularize the importance of these norms in the early stages of a researcher's journey into the publishing world. During the formative years in the field, early-career researchers might be naturally drawn to emulating the behaviour of the more experienced researchers in their labs and workplaces. An educational push for holding research ethics, norms, and standards as the paramount element of doing research and development in a consequential field like AI needs to be strongly encouraged. Principal Investigators (PIs) and other senior researchers in both academic and industry labs should shepherd those who are still growing accustomed to the modes of operation of research in the domain.

---

[19] Brown, T. (2008). Design thinking. Harvard business review, 86(6), 84.
[20] *NeurIPS 2020 Call for Papers*. (n.d.). NeurIPS. Retrieved from: https://nips.cc/Conferences/2020/CallForPapers
[21] ICLR Code of Ethics. Ninth International Conference on Learning Representations. Retrieved from: https://iclr.cc/public/CodeOfEthics

The widespread presence of such a requirement could also incite greater awareness among technical researchers in the field of AI regarding fields and areas like ethics, psychology, anthropology, sociology, and critical race theory. Furthermore, making impact statements and other educational pushes necessary could spark interdisciplinary collaborations between researchers.

### 5. Require the publication of expected results

Requiring that researchers write and publish the expected results of their research project (including but not limited to its broader social and ethical impacts) could help foster reflection around potential benefits and harms even before researchers undertake their project. For example, a list of journalists who would be interested in hearing about the benefits of the author's thesis on the rest of society could be created, incentivizing the consideration of their project benefits (scientists will want their papers to be promoted in newspaper articles). This would not only encourage researchers to reflect on the impacts of their work, but perhaps guide them towards generally less risky and more beneficial research. This could be similar to how many psychologists now publish a preliminary report on how they will conduct their research, the variables they will be measuring, and their hypothesis, before they perform an experiment. This prevents data manipulation in attempts to find statistically significant results.[22] In the language of software engineering, it could create something akin to Test-Driven Development (TDD) whereby the researchers can preemptively highlight what it is that they are aiming to achieve with their research and view negative results in a positive light as a way of showing ways that don't lead to results that were stated as the goals of the research nonetheless led to an understanding of some aspects of the field helping to improve the knowledge base for future researchers in the domain.

Of course, having much greater diversity of researchers in the field of AI—in terms of race, gender identity, sexual orientation, geography, language, lived experiences, and socioeconomic status—could also make a significant difference in how and how much researchers think about the impacts of their work. It seems reasonable that, in at least some cases (if not many), the potential benefits and pitfalls of researchers' work may be more pronounced for a certain demographic, and people who are part of that demographic are more likely to be attuned to how it may impact them. Thus, having more diverse actors in the field of AI could help foster more comprehensive reflection on the potential impacts of a research project. Additionally, a framing whereby the researchers take on the onus to surface these impacts rather than placing the burden on the people who might be impacted (often they might not have adequate knowledge, resources, or abilities) to defend themselves creates a more pro-social way forward for conducting research.

---

[22] Kim, A. B. (November 1, 2019). *Psychologists confront impossible finding, triggering a revolution in the field*. CBC News. Retrieved from:
https://www.cbc.ca/listen/live-radio/1-23-ideas/clip/15744568-psychologists-confront-impossible-finding-triggering-revolution-field?onboarding=false

### 6. Revamp the peer review process

The well-established practice of peer review is a great opportunity for exchanges on the risks and benefits each reviewer sees in the paper they are revising. If a question or requirement were added to this effect when papers were reviewed, it may have a rapid and widespread impact in inciting researchers to consider what may follow from their research. This could be done in a way that is similar to how the World Health Organization (WHO) highlights Dual Use Research of Concern (DURC): research in the realm of the life sciences "that is intended for benefit, but which might easily be misapplied to do harm."[23] To have a similar category for AI research would be pertinent, and developing guidelines for this kind of research seems crucial, both to define it and to control it, especially given that AI squarely falls under the category of a dual-use and general-purpose technology. Such guidelines might include needing special authorization to conduct this type of research, or it may be required that the research be reviewed more extensively before being published.

Further, it's obvious that an effective review process should promote limiting risks while also being clear, fair, and efficient. One way of doing this is to intensify the requirements for publication proportionally to how risky the research is deemed by peers reviewing the paper. One way to go about this could be to identify a list of specific risks that any AI model proposed in a research paper could have. These risks may have numerical weights assigned to them depending on how dangerous their consequences may become. Each peer reviewer could identify the risks they deem relevant to the AI model they are examining, and assign a score between 0 and 5 to each risk in accordance with how likely it is that the risk will materialize. The risks' weights could then be multiplied by their corresponding likelihood score. Each of these quotients can be compiled into a sum. The sums obtained by each reviewer for one paper can then be added up and made into an average. If this 'total risk average' is higher than a pre-established number, the paper could then be immediately rejected, require further reviewing and discussion by a third-party group of experts, or be published under much stricter requirements than other papers that are below the 'total risk average' threshold. Such a threat modelling approach is already used extensively in the field of cybersecurity to prioritize risks and vulnerabilities to guide the efforts of researchers and practitioners in working to address them[24].

This process, or one with a similar structure, has the advantage of remaining somewhat efficient by not requiring that papers that are considered non-risky be subjected to more time-consuming reviewing or more stringent requirements unnecessarily. Of course, a procedure like the one presented above risks potentially wrongly identifying a model as low risk, whereas a model that held all papers to higher standard regarding risk avoids this. However, it's unlikely that applying more stringent standards to all papers is necessary, efficient, or realistic considering the sheer volume of papers published. A more targeted approach seems better suited to the reality of publishing in the domain of AI. A consideration that publishers can adopt is to think about the rate of false positives and track them over the period of reviews and adjust as they go along to judge the efficacy of this mechanism.

---

[23] *WHO | Dual Use Research of Concern (DURC)*. (n.d.). World Health Organization. Retrieved from: https://www.who.int/csr/durc/en/

[24] Bodeau, D., McCollum, C., and Fox, D. (2018). Cyber threat modeling: Survey, assessment, and representative framework. The Homeland Security Systems Engineering and Development Institute, Tech. Rep.

To the risk of this becoming repetitive, the cornerstone of an effective review process, as previously mentioned, will be a diverse body of reviewers from the point of view of race, gender, disability, socioeconomic status, geography, and more. Risk assessment is unlikely to be fair and comprehensive if all reviewers have similar backgrounds and experiences. This will result in direct harm to those not represented among reviewers, which are often individuals who are already marginalized and most at risk. This is something that needs to be created proactively and will not necessarily emerge organically since reviewers are often sourced from a tightly knit network of people one is already familiar with; breaking free from that requires constant and conscious effort.

## Potential Drawbacks

1. **Constrained innovation**

One important area where there may be some pitfalls is innovation. If publishing norms are more stringent, then cutting-edge research may be ignored or underfunded. In some cases this may be because the potential risks outweigh the benefits. However, in other cases it may be the result of a lack of awareness from the researcher's part on what those publication norms are. This could have a particularly negative impact on emerging scholars in AI from regions and countries where AI research and development is in nascent stages, especially those which have a less than proportionate representation in the scientific publications at major conferences and journals (which are mostly concentrated in the Western hemisphere).[25] This could harm scientific diversity in the field.

One can undoubtedly see this as a missed opportunity, but it's better framed as an opportunity for better work and innovation. What if we could use this as an opportunity to build AI that did more good than harm? By erring on the side of caution, we may encourage researchers to better understand the consequences of developing and deploying a certain system or application. Understanding the tail risks of innovation is critical. The insights from fields such as cryptography are that the risks are enormous from poor research, and given the nature of AI systems, a similar level of risk is to be expected. In light of this understanding of asymmetric returns, we need to raise our standards of what constitutes "innovation," and who gets to decide if a new technological application constitutes positive innovation; bringing more good than harm. From this perspective, constraining innovation is not a pitfall of changing publishing norms. Instead, changing publishing norms may foster higher quality, more inclusive and positive innovation. It can be viewed as a mechanism to bend the field of research towards a prosocial direction, something that governments use frequently in the form of regulations to guide how innovation happens in the market.

---

[25] Saurabh, Mishra; Perrault, Raymond; Shoham, Yoav; Brynjolfsson, Erik; Etchemendy, John; Grosz, Barbara; Lyons, Terah; Manyika, James; Niebles, Juan C. (2019). The AI Index 2019 Annual Report. Human-Centered Artificial Intelligence Institute. Stanford University. Retrieved from: https://hai.stanford.edu/sites/default/files/ai_index_2019_report.pdf

In a similar vein, like many other scientific disciplines, machine learning and artificial intelligence have been affected by their own reproducibility crisis, where a worrying number of algorithmic research results cannot be reproduced when other data scientists run the same experiments.[26] This is of particular concern in fields such as digital biomedicine, where faulty models for disease diagnosis and monitoring can place human lives at great risk. One key component of the problem is the absence of information about the training and evaluation code, the number of training runs required, and datasets used.[27] Another part of the issue is the environment in which data scientists operate, where there is a pressure to publish quickly, a reluctance to report failed replications and a lack of computational and human resources to test every condition and fine-tune each hyperparameter.[28] Better publication norms will not only encourage more rigour in scientific methodology, but also limit the number of cases where research is modelled on "false starts." The danger of this can ensure that future innovation in the field is based on a set of verifiable and veritable discoveries. In an inherently stochastic domain like AI, the degree of variability that can occur in experiments where one might possibly find an experimental run through sheer luck to get the results that showcase a correlation that they want is all the more reason for advocating for higher rigour; a focus on causation over correlation[29] is going to be essential in addition to the points mentioned above.

2. A "black market" for research

With more stringent norms on what gets published in the field of AI, there is a possibility that research deemed too risky will be driven underground. Meaning, researchers whose work is rejected may publicize their research in other ways, circumventing the measures put in place. Due to the stricter publishing norms, the rejected research may become even more dangerous as it's excluded from mainstream critique and necessary scrutiny. Put simply, stringent publishing norms may actually increase the likelihood of harmful AI by creating a "black market" for research that has subversive aims.

---

[26] Hutson, Matthew. (2018). *Artificial intelligence faces reproducibility crisis*. Science Magazine. Retrieved from: https://science.sciencemag.org/content/359/6377/725

[27] Stupple, Aaron; Singerman, David; Celi, Leo A. (2019). The reproducibility crisis in the age of digital medicine. *Npj | Digital Medicine, (2)2,* 1-3. Retrieved from: https://www.nature.com/articles/s41746-019-0079-z

[28] Pineau, Joelle; Vincent-Lamarre, Philippe; Sinha, Koustuv; Larivière, Vincent; Beygelzimer, Alina; d'Alché-Buc, Florence; Fox, Emily; Larochelle, Hugo. (2020). Improving Reproducibility in Machine Learning Research (A Report from the NeurIPS 2019 Reproducibility Program). https://arxiv.org/abs/2003.12206

[29] Pearl, J. (2009). *Causality*. Cambridge University Press. Retrieved from: https://www.cambridge.org/core/books/causality/B0046844FAE10CBF274D4ACBDAEB5F5B

Of course, there are counterarguments against this scenario. First, the field of AI/ML is heavily dominated by individuals who are affiliated with universities and companies. In both cases, getting research published by a reputable journal or presented at a recognized conference is key to advancing their career. Simply putting out research without it being published or affiliated with their institution (whether academic or corporate) because it is too risky according to publishing norms is of limited use. Hence it seems unlikely that someone with the knowledge and qualifications necessary to build innovative yet risky AI would have an incentive to share it using an alternative method.

There may be a greater incentive for scientists to conduct research that is more likely to be published in accordance with the new, more stringent standards. It is highly unlikely that the risks for underground research are big enough to warrant *not* moving towards more stringent publishing standards in terms of security and risk in the field of AI. It's also important to note that the field's current exclusivity to those with affiliations to large and wealthy institutions is not in itself positive. We believe the field of AI should be more accessible than it currently is in that regard.[30]

### 3. Misplaced accountability

Paradoxically, there are questions as to whether it's actually more dangerous if risky AI research goes "underground" than if it's publicized and sanctioned by highly regarded institutions and publications. If an AI research submission is rejected by a publication, then it's likely that it'll receive minimal attention. However, if risky AI research were to be published by these journals and publications then it's likely it may do significant harm because people (especially academics and fellow researchers) tend to refer to these sources for the latest relevant information about progress in the field; under the assumption that the information published has been peer-reviewed. It's this same research that often gets widespread media coverage as well. This follows from the principle of using "strategic silence" as a way of limiting the *oxygen* that is provided to mis-, dis-, and mal-information by not offering it a platform by way of discussion, time, and resources.[31]

---

[30] Gupta, Abhishek; Lanteigne, Camylle; Kingsley, Sara. (2020). SECure: A Social and Environmental Certificate for AI Systems. Montreal AI Ethics Institute. Retrieved from: https://arxiv.org/abs/2006.06217
[31] Wardle, Claire; Derakhshan, Hossein. (2017). Information disorder: Toward an interdisciplinary framework for research and policy making. Council of Europe. Retrieved from: https://rm.coe.int/information-disorder-toward-an-interdisciplinary-framework-for-researc/168076277c

Furthermore, general audiences are less likely to scrutinize research sanctioned by a journal or an institution. In other words, when risky papers are published, we may not be able to count on public scrutiny to highlight its dangerous possibilities. There have been numerous instances of popular media coverage for dubious research stemming from low-quality journals, preprint servers, and other places that had a significant impact on the public's perception of AI. As an example, research that claimed to create a "gaydar" identifying a person's sexual orientation was thoroughly debunked by leading researchers in the field of AI,[32] but the harmful and offensive work still received significant media attention.[33][34] We posit that this might be because of the highly technical nature of the field (not that this isn't the case in other domains, but there is a disproportionate attention paid to the advances in this field while the ability of the general public to parse the advances for what they truly are might not be sufficient) which leads to an overestimation in the capabilities of the systems.

The fact that such research was published—and thus, endorsed by the publication or institution behind it—created a false sense of security and legitimacy for many people. Papers that are published are usually revised and reviewed, and this process, along with the metaphorical seal of approval from the publication/institution, can obfuscate the overly risky nature of the results being published. It essentially shields the paper from any scrutiny because, at first glance, the paper has all the characteristics of an acceptable or even outstanding paper. Thus, getting widespread public scrutiny, especially from individuals outside the AI community, is likely to be quite difficult as nothing seems particularly questionable.

With this in mind, there is a responsibility on the part of the journal's editorial board to have experts outside of the field review papers in order to identify errors or risks with the research. Homogenous editorial boards (i.e. dominant avatars of society), don't have the required perspective to ask essential, critical questions like that of a diverse board. In this sense, a diverse board will be able to provide additional points of failure for controversial scrutiny, before even requiring any public intervention. Hence, stricter publication norms that include such a board would be desirable. In addition, one of the mechanisms trialled with NeurIPS and ICML over the last few months is the concept of ML Retrospectives[35] (co-organized by Abhishek Gupta, one of the authors of this document) where researchers are asked to reflect on their prior work and identify shortcomings, improvements, changes, and any other comments that they have on those papers to highlight the need to revise understandings of discoveries over time as new evidence and knowledge comes to light. Normalizing the "owning" of faults in prior publications can move accountability back to authors, no matter their prestige, leading to a healthier intellectual ecosystem.

---

[32] y Arcas, Blaise; Mitchell, Margaret; Todorov, Alexander. (2017). *Physiognomy's New Clothes*. Medium. Retrieved from: https://medium.com/@blaisea/physiognomys-new-clothes-f2d4b59fdd6a

[33] Burdick, Alan. (2017). *The A.I. "Gaydar" Study and the Real Dangers of Big Data*. The New Yorker. Retrieved from:
https://www.newyorker.com/news/daily-comment/the-ai-gaydar-study-and-the-real-dangers-of-big-data

[34] Vincent, James. (2017). *The invention of AI 'gaydar' could be the start of something much worse*. The Verge. Retrieved from:
https://www.theverge.com/2017/9/21/16332760/ai-sexuality-gaydar-photo-physiognomy

[35] Lowe, Ryan. (2019). *Introducing Retrospectives: 'Real Talk' for your Past Papers*. The Gradient. Retrieved from: https://thegradient.pub/introducing-retrospectives/

4. **Reduced pace of publication**

Research progress may be slowed to allow for more thorough risk evaluation. This may affect the quick rate at which AI papers are currently published. Nevertheless, pausing the influx of information on AI can provide more time to dissect, digest, and process research. Furthermore, a stricter process to pass, and one that takes time, could discourage researchers against cutting corners, which could lead to having their paper rejected and making the process unnecessarily long. This can then help guard against unnecessary research being proposed. One of the trends of taking rejected submissions from one conference or journal and making quick, minor tweaks and pushing it out to the next conference is a particularly problematic scenario which encourages the behaviour whereby researchers attempt a "keep trying to till you win" mindset that has the potential to lower the quality of overall scholarship in the field. Making explicit perhaps where this idea was submitted before and reasons for rejections, how those shortcomings were addressed can serve two purposes: improve allocation of scarce reviewing resources in the field and push higher standards of transparency in the field. Cross-linking submissions on a platform like OpenReview (acknowledging though that there are many caveats on the platform) can help us move in that direction.

5. **Lost control by publishers**

Some publishers may feel as if they have lost an element of control over what they publish if more stringent norms are put into place. They may also perceive these new norms as constricting what they traditionally viewed as "worthy" of scientific regard. Whether true or not, many publications perceive themselves as being apolitical; taking an explicitly agnostic stance towards scientific developments. Therefore, norms that have an arguably social and/or political bent could present a fundamental challenge and potentially give rise to an increase in the bifurcation of journals split on social, political, or other lines.

6. **Existing power structures and inequities in AI**

There are a host of concerns that must be accounted for in the case that an external body—such as a journal, a norm-enforcement committee, an ethics board, etc.—oversees the release of research findings; one being the importance of equal opportunity and diverse representation in the technological sphere. This standard not only ensures that the domain of technological research remains democratic, but also prevents it from becoming an echo chamber or a site of informational homogeneity.

In order to frame best publication practices for high-stake AI-related research, we must ensure that this set of practices is predicated upon a firm anti-oppression mandate, whether this oppression is related to socioeconomic status, race, gender, ethnicity, mental/physical disability, or any other characteristic that incites unjust discrimination. Responsible publication norms, by their very nature, must not only strive to mitigate harms related to the exposure of certain research findings, but also to inhibit the *epistemic threat* of information suppression under the guise of best practices.

In order for a novel set of publication practices to be truly responsible in nature, it must be acknowledged that while a regulatory framework for AI research has the potential to reduce threats to public safety, it may also inadvertently serve as a site of gatekeeping, suppression and other forms of exclusionary practice. Such a framework therefore risks consolidating epistemic injustice within the academic, technological and political spheres of influence, the prevention of which will require a defined set of safeguarding measures. Injustice can be mitigated through the use of radical transparency where the publisher maintains an open list of papers that were rejected and the reasons for rejection in a public repository that is subject to public scrutiny and analysis preventing biases against any groups.

To ensure that reactive measures for harm mitigation (i.e. flagging papers post-publication) are not discriminatory in nature or grounded in any form of identity-based prejudice, a more robust set of flagging requirements must be established. In addition to allowing the user to select a general reason for flagging certain content, the process could be made less enticing and/or gameable to those with unfounded reasoning and discriminatory motives by requiring a more exhaustive and substantiated explanation as to why the content should be removed. To this end, making the process involved enough to prevent trolling behaviour can be one proposed approach.

In the case of suspected discriminatory practice throughout the publication process, there ought to be an appeal protocol in place whereby subjects of discrimination can formalize a complaint against the relevant publication or platform and negotiate reconsideration of their research by a designated secondary screening committee. Such a protocol may include an instrument of a similar nature to that of the checklist or DREAD-like (Damage, Reproducibility, Exploitability, Affected users, Discoverability) scoring principles[36]. Rather than measuring security threats, this instrument could assess and systematize the long-term personal impacts of discriminatory publication practices whether they be social, professional, or economic in nature. Within a novel regulatory framework for responsible release norms, this impact checklist could be appealed to and taken up with a third party entity in instances of publication prejudice; this entity having authority over the reconsideration of initially rejected research, along with the pursuit of certain disciplinary measures if necessary. Such an organizational framework could serve to encourage publisher accountability, caution against future discriminatory practices, and ultimately preserve epistemic justice in the domain of AI research.

---

[36] Shostack, A. (2008). Experiences Threat Modeling at Microsoft. MODSEC@ MoDELS. Retrieved from: https://www.researchgate.net/publication/228793485_Experiences_Threat_Modeling_at_Microsoft

### 7. Malicious applications of alternative release norms

More staged or closed publication practices pose a threat to democratized exposure under the guise of responsible release norms, which may result in the omission of crucial information or gatekeeping. For instance, if research unveils potential risk of a given AI/ML model and there exists a conflict of interest between the publication platform and stakeholders invested in the model, the pretence of "alternative release strategies" could serve as something of a loophole for these stakeholders to suppress certain information and preserve their interests (whether financial, social, etc.) without being held liable for censorship.

The malicious application of release norms not only entails self-interested nondisclosure, but also the suppression of projects carried out by actors who occupy marginalized identities, which could theoretically be falsely attributed to the overstated "harms" that their findings may yield, should they be disclosed publicly. Such scenarios lend themselves to the risk of epistemic discrimination against marginalized groups. They can also have a chilling effect on the field of research and promote underground attempts at recreating the systems without proper controls in place because of this obfuscation of the real impacts of the system, thus potentially leading to more harm.

## Conclusion: Ways Forward

### 1. State clearly and consistently the need for established norms

It's clear that additional restrictions and/or guidelines in the field of AI are required because there are risks that need to be addressed without exception. Unfortunately, however, there is no consensus on this in the field. Therefore, more individual researchers, institutions, governments, etc. need to clearly and publicly communicate the potential societal impacts of AI research in order to convey the need for third party regulation. In addition, more researchers and institutions supporting and replicating efforts like the NeurIPS impact statement requirement would make such considerations more standardized across the field.[37]

### 2. Coordinate and build trust as a community

Forming alliances, making connections, and building relationships are the hallmarks of effective community collaboration. One way this can be enacted is through connecting with a publication standard-setting body, whereby concerned researchers and institutions would be able to lobby for greater action on the topic and witness first-hand how the publication norms are created. For example, a potential partnership with COPE[38] could prove fruitful. COPE is an organization committed to educating and supporting editors, publishers and those involved in publication ethics with the aim of moving the culture of publishing towards one where ethical practices become a normal part of the publishing culture. So far, it seems they haven't done any work related to the field of AI, but perhaps they would be interested in expanding their work to account for new ethical issues arising in the technology and computer science space.

---

[37] Ashurst, Carolyn; Anderljung, Markus; Prunkl, Carina; Leike, Jan; Gal, Yarin; Shevlane, Toby; Dafoe, Allan. (2020). *A Guide to Writing the NeurIPS Impact Statement*. Centre for the Governance of AI. Retrieved from:
https://medium.com/@GovAI/a-guide-to-writing-the-neurips-impact-statement-4293b723f832

On a larger scale, a solid foundation for this collaboration within the community is the potential for an institution or process that is recognized worldwide and respected in terms of trust around technology. MAIEI's work on SECure[39] and Green Lighting ML[40] are potential options. These initiatives could provide a standardized comparative measure to evaluate the trust in AI systems across the broad range of topics and themes contained within the community, providing a much-needed *lingua franca*.

### 3. Change the approach

As previously mentioned with COPE, the field of AI must work towards a new publication culture that prioritizes ethical practices and de-prioritizes progress for the sake of progress. To create safe and responsible AI, the core values and culture of the field must be informed by ethical practices, principles, and guidelines. Additionally, it may be important to place a greater emphasis on the application of the technology after it has been deployed, rather than simply the publishing of the information itself. For example, Canada has developed National AI Standards, which regulate the application of the technology *rather* than the published content about said technology.[41] This isn't to say that AI publication norms shouldn't be created and followed, but that they should be done in parallel with other efforts necessary to ensure that the science and technology of AI are developed in a more humane manner.

---

[38] *About COPE*. Committee on Publication Ethics (COPE). Retrieved from: https://publicationethics.org/about/our-organisation

[39] Gupta, A., Lanteigne, C., and Kingsley, S. (2020). SECure: A Social and Environmental Certificate for AI Systems. arXiv preprint. Retrieved from: https://arxiv.org/abs/2006.06217

[40] Gupta, A., and Galinkin, E. (2020). Green Lighting ML: Confidentiality, Integrity, and Availability of Machine Learning Systems in Deployment. arXiv preprint. Retrieved from: https://arxiv.org/abs/2007.04693

[41] Coop, Alex. (2019). *Canada's CIO Strategy Council publishes national AI standards*. IT World Canada. Retrieved from: https://www.itworldcanada.com/article/canadas-cio-strategy-council-publishes-national-ai-standards/422722